\documentclass[english,aps,prl,twocolumn,amsmath,amssymb,showpacs,superscriptaddress,notitlepage,longbibliography]{revtex4-2}
\usepackage[T1]{fontenc}
\usepackage[latin9]{inputenc}
\usepackage{color}
\usepackage{dsfont}
\usepackage{amsmath}
\usepackage{amssymb}
\usepackage{graphicx}
\usepackage{esint}

\makeatletter
\usepackage{amssymb}
\usepackage{amsmath}
\usepackage{graphicx}
\usepackage[colorlinks=true,linkcolor=blue,anchorcolor=red,citecolor=blue,urlcolor=blue]{hyperref}
\usepackage[caption=false]{subfig}
\usepackage{lipsum}

\makeatother

\usepackage{babel}
\begin{document}
\renewcommand{\figurename}{Fig.}
\title{Exceptional Non-Hermitian Topology Associated with Non-Toroidal Brillouin Zones}
\author{W.\ B. Rui}
\email{wbrui@hku.hk}

\address{Department of Physics and HK Institute of Quantum Science \& Technology,
	The University of Hong Kong, Pokfulam Road, Hong Kong, China}

\author{Z.\ D. Wang}
\email{zwang@hku.hk}
\address{Department of Physics and HK Institute of Quantum Science \& Technology,
	The University of Hong Kong, Pokfulam Road, Hong Kong, China}
\address{Hong Kong Branch for Quantum Science Center of Guangdong-Hong Kong-Macau Great Bay Area, Shenzhen, China}
\date{\today}
\begin{abstract}
Exceptional points (EPs) are prominent non-Hermitian band degeneracies
that give rise to a variety of intriguing and unconventional phenomena. 
Similar to Weyl and Dirac points, EPs carry topological charges
and comply with the celebrated fermion doubling theorems in lattices.
Beyond these characteristics, EPs exhibit more exotic topological 
properties, particularly non-Abelian braiding topologies not seen in conventional degeneracies.
However, these core concepts of EPs have been
established under the assumption of toroidal Brillouin zones. Here, we investigate EPs 
in two-dimensional non-Hermitian lattices where the 
fundamental domain of the Brillouin zone is a Klein bottle, rather 
than a torus assumed in previous studies. We find that EPs 
do not necessarily appear in pairs with opposite topological charges 
in the Brillouin Klein bottle, thus violating the fermion doubling theorem. 
The violation occurs because, without crossing the boundary, the sum of 
the topological charges of EPs is in fact an even number rather than zero. Moreover,
we uncover unique braiding topologies of EPs that cannot be captured
by existing theories. Specifically, the composite braidings
around all EPs equals the braiding along the boundary of the Brillouin
Klein bottle. This novel braiding topology further confirms the failure
of the fermion doubling theorem, and allows us to explore the non-Abelian
braidings of EPs beyond the scope of topological charges. Our work
highlights the fundamental role of Brillouin-zone topology in non-Hermitian
systems.
\end{abstract}
\maketitle
\textit{\textcolor{blue}{Introduction.}}\textit{\textemdash{}} Exceptional
points (EPs) represent one of the most intriguing aspects of non-Hermitian
physics, drawing significant attention across diverse fields such
as optics, photonics, acoustics, electronics, and condensed matters~\citep{miri_exceptional_2019,ozdemir2019paritytextendashtime,gao2015observation,Schindler-EP-electronics,li2023exceptional,feng2014singlemode,hodaei2014paritytimetextendashsymmetric,peng2014lossinduced,wiersig2014enhancing,chen2017exceptional,hodaei2017enhanced,liu2016metrology,Guo-unidirectional-2009,lin2011unidirectional,regensburger2012paritytextendashtime,peng2014paritytextendashtimesymmetric}.
At EPs, both eigenvalues and eigenvectors coalesce~\citep{katoPerturbationTheor1995,berry_physics_2004,heiss_physics_2012},
leading to a variety of unconventional physical phenomena, including
enhanced sensitivity~\citep{wiersig2014enhancing,chen2017exceptional,hodaei2017enhanced,liu2016metrology}
and unidirectional invisibility~\citep{Guo-unidirectional-2009,lin2011unidirectional,regensburger2012paritytextendashtime,peng2014paritytextendashtimesymmetric}.
In the rapidly growing field of non-Hermitian topology~\citep{Leykam_PRL_2017,Xu_weyl_nh,shen_PRL_2018,zhou_observation_2018,Wang_PRL_2018,Bergholtz_PRL_2018,Zhang-correspondence,Okuma-origin,Rui_dirac_nh,Gong-topology,Kawabata-symmetry,li2020critical,Rui_PhysRevB20192,Rui_Hermitian_2023,ding2022nonhermitian,Rui_PhysRevLett_2023,Hu_PhysRevLett_2021,Shuichi-Non-Bloch,helbig2020generalized,xiao2020nonhermitian,ashida2020nonhermitian},
the importance of EPs has been further highlighted. These unique non-Hermitian
degeneracies serve as defining characteristics of a large class of
non-Hermitian topological semimetals, known as exceptional semimetals~\citep{Kawabata_ep_2019,Bergholtz-Exceptional,Carlstr-Exceptional-links,Yoshida-PRB-2019,rui-susy,Jin-PRL-2019,Yang-prl-2020,tang2020exceptional,Mandal-prl-2021,Liu-PRL-2021,rui-nh-fermi-double,tang2023realization,wu2024thirdorder}.

Like Weyl and Dirac points, EPs carry topological charges and are
subject to the celebrated fermion doubling theorem, a universal no-go
theorem governing both Hermitian and non-Hermitian lattice systems~\citep{nielsen_no-go_1981,nielsen_absence_1981,nielsen_absence_1981-1,Chiu-NH_Double}.
This theorem dictates that EPs must appear in pairs with opposite
topological charges in lattices. The underlying proof relies on the
periodic boundary conditions of the Brillouin torus~\citep{Chiu-NH_Double},
which ensure that the total topological charge of EPs\textemdash equal
to a line integral along the torus boundary\textemdash sums to zero.

While EPs share these fundamental features with conventional band
degeneracies, they also exhibit unique and extraordinary topological
characteristics, particularly non-Abelian braiding topologies~\citep{Zhi-Homotopical,Hu_ep_knot,Hu_Exceptional_Non-Abelian,Wojcik_eigenvalue,Wojcik_Homotopy,patil2022measuring,Lukas_PhysRevResearch,wang2021topological,Qiu-PRL-2023}.
These braiding topologies arise because the complex eigenenergies
in the vicinity of EPs can braid around each other, and become non-Abelian
when there are more than two eigenenergies. The braiding topology
of EPs in lattices follows specific rules~\citep{Wojcik_eigenvalue,Hu_ep_knot}:
the composite braidings around all EPs must match the braiding along
the boundary of the Brillouin torus.

These foundational concepts serve as the cornerstone for studying
EPs in lattices. However, they have been developed under the assumption
that the Brillouin zone is topologically a torus. Recent discoveries
have shown that, under momentum-space nonsymmorphic symmetries, the
fundamental domain of the Brillouin zone can adopt a non-toroidal form, 
specifically, it can form a Klein bottle rather than a torus ~\citep{chen2022brillouin,Zhao_Momentum-Space,chen2023classification,Fonseca_PRL_2024,Yang-Nonsymmorphic,Li-klein-prb,Lai_Real-projective-plane,Tao_PRB_2024}.
This raises an important question: Do the fermion doubling theorem
and the established braiding topology of EPs, both of which depend
on torus boundary conditions, still hold true in the Brillouin Klein
bottle?

In this work, we demonstrate that the fermion doubling theorem breaks
down, and a unique braiding topology of EPs emerges in the non-Hermitian
Brillouin Klein bottle. We focus on the region without crossing the
Klein bottle boundary, denoted as $K^{2}$, where a local orientation
can be defined to avoid the sign ambiguity of topological charges
due to the global non-orientability of the Klein bottle. In this region,
the topological charges of EPs maintain definite signs and behave
similarly to conventional ones. We find that the sum of all topological
charges within $K^{2}$ equals a line integral along the Klein bottle boundary
$\partial K^{2}$, and satisfies
\begin{equation}
	\ensuremath{\sum_{\mathbf{k}_{i}\in K^{2}}v(\mathbf{k}_{i})=\oint_{\partial K^{2}}d\mathbf{k}\cdot\nabla_{\mathbf{k}}\log\Delta(\mathbf{k})\in2\mathbb{Z}},\label{eq:K2-rule}
\end{equation}
where $v(\mathbf{k}_{i})$ is the topological charge of the EP located
at $\mathbf{k}_{i}$, and $\Delta(\mathbf{k})$ is the discriminant
defined below Eq.~(\ref{eq:discriminant-number}). Since the total
charge is an even number rather than zero, EPs do not necessarily
appear in pairs with opposite charges, leading to the breakdown of
the fermion doubling theorem. Furthermore, we establish the braiding
topology of EPs in the Brillouin Klein bottle, and find that the previously
established braiding rules do not apply. Instead, the composite of
the braidings around all EPs in $K^{2}$, denoted as ($\mathsf{b}_{1}\mathsf{b}_{2}\cdots\mathsf{b}_{n}$), equals to the braiding along
the boundary $\partial K^{2}$ , expressed as ( $\mathsf{b}_{a}\mathsf{b}_{b}\mathsf{b}_{a}\mathsf{b}_{b}^{-1}$). That is,
\begin{equation}
	\mathsf{b}_{1}\mathsf{b}_{2}\cdots\mathsf{b}_{n}=\mathsf{b}_{a}\mathsf{b}_{b}\mathsf{b}_{a}\mathsf{b}_{b}^{-1},
\end{equation}
which can further confirm the breakdown
of the fermion doubling theorem. Finally, we explore the non-Abelian
braidings associated with EPs, revealing rich topologies that cannot
be characterized by topological charges alone.

\begin{figure}[t]
\includegraphics[width=0.95\columnwidth]{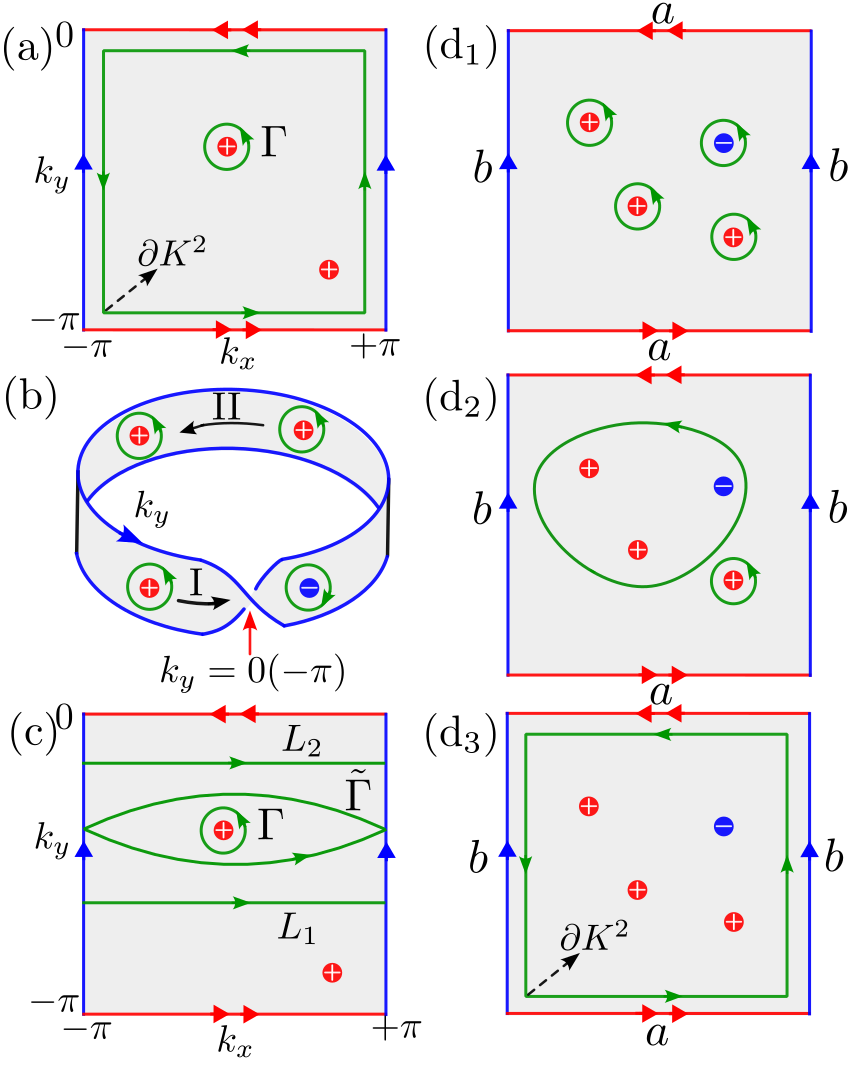}\caption{(a) Depiction of the Brillouin Klein Bottle, showing EPs (red and
blue dots) and the oriented loop $\Gamma$ used to define the topological
charge. (b) Illustration of the global non-orientability of the Brillouin
Klein Bottle, where the twist is introduced after gluing the red edges
in (a). (c) In the gray region, the loop $\Gamma$ can be continuously
deformed to two lines $L_{1}$ and $L_{2}$. (d$_{1}$-d$_{3}$) represent
the continuous deformation of integration paths to the boundary of
Brillouin Klein bottle.\label{fig1}}
\end{figure}

\textit{\textcolor{blue}{Brillouin Klein Bottle in non-Hermitian systems.}}\textit{\textemdash{}}
Let us consider a non-Hermitian Hamiltonian $\mathcal{H}(\mathbf{k})$
in two dimensions that respects the momentum-space glide reflection
symmetry as~\citep{chen2022brillouin} 
\begin{equation}
U\mathcal{H}(k_{x},k_{y})U^{-1}=\mathcal{H}(-k_{x},k_{y}+\pi),\label{eq:k-glide-sym}
\end{equation}
where $U$ is a unitary operator. The symmetry maps $(k_{x},k_{y})$
to ($-k_{x},k_{y}+\pi)$, partitioning the first Brillouin zone into
two equivalent regions: $(k_{x},k_{y})\in[-\pi,\pi)\times[-\pi,0)$
and $[-\pi,\pi)\times[0,\pi)$. Hence, it suffices to consider one
of these regions, e.g., $[-\pi,\pi)\times[-\pi,0)$ as plotted in
Fig\@.~\ref{fig1}(a). 

Examining the two horizontal edges (red lines) of this region at $k_{y}=-\pi$
and $0$, we find that they must be glued together in opposite directions
due to the symmetry $U\mathcal{H}(k_{x},-\pi)U^{-1}=\mathcal{H}(-k_{x},0)$.
In contrast, the two vertical edges (blue lines) should be glued together
in the same direction. This specific edge identification results in
a Klein bottle rather than a torus, which is termed the Brillouin
Klein bottle.

\textit{\textcolor{blue}{EPs in Brillouin Klein Bottle.}}\textit{\textemdash{}}
EPs carry topological charges known as discriminant numbers, which
are defined as~\citep{shen_PRL_2018,Chiu-NH_Double} ,
\begin{equation}
v(\mathbf{k}_{i})=\frac{i}{2\pi}\oint_{\Gamma(\mathbf{k}_{i})}d\mathbf{k}\cdot\nabla_{\mathbf{k}}\log\Delta(\mathbf{k}).\label{eq:discriminant-number}
\end{equation}
where $\Delta(\mathbf{k})=\Pi_{j<k}[E_{j}(\mathbf{k})-E_{k}(\mathbf{k})]^{2}$
is the discriminant, and $E_{j}(\mathbf{k})$ is the $j$th eigenvalue
of $\mathcal{H}(\mathbf{k}).$ It is important to emphasize that $\Gamma(\mathbf{k}_{i})$
is a small loop that encircles the EP at $\mathbf{k}_{i}$ in a \textit{counterclockwise}
orientation.

One might try to directly use the discriminant number~(\ref{eq:discriminant-number})
for EPs in the Brillouin Klein bottle, as shown in Fig\@.~\ref{fig1}(a).
However, a fundamental challenge arises: the Klein bottle is globally
\textit{non-orientable}, meaning that a consistent counterclockwise
or clockwise orientation cannot be maintained over the entire surface.
As highlighted by process \textquotedbl I\textquotedbl{} in Fig\@.~\ref{fig1}(b),
when an EP and its associated loop traverse the twist at $k_{y}=0\,(-\pi)$
(red arrow) and return to its original position, the counterclockwise
orientation is reversed to a clockwise orientation. Consequently,
the sign of the invariant becomes ambiguous; i.e., $v(\mathbf{k}_{i})$
cannot be distinguished from $-v(\mathbf{k}_{i})$ over the entire
surface. This phenomenon is also observed for the Chern number in
Brillouin Klein bottles~\citep{chen2022brillouin,Fonseca_PRL_2024}.
Therefore, the global non-orientability poses a challenge for verifying
the fermion doubling theorem regarding the pairing of EPs with opposite
topological charges.

\textit{\textcolor{blue}{Local versus global orientability.}}\textit{\textemdash{}}
In non-orientable manifolds, while a global orientation cannot be
established, it is actually feasible to define a local orientation~\citep{allen,Fonseca_PRL_2024}.
Here, we focus on the region without crossing the Klein bottle boundary,
denoted as $K^{2}$ as shown by the gray area in Fig\@.~\ref{fig1},
so that the the twist at $k_{y}=0\,(-\pi)$ can be avoided. Within
this region, a counterclockwise orientation can still be meaningfully
defined, and the sign of topological charges does not change when
moving EPs, as illustrated by process \textquotedbl II\textquotedbl{}
in Fig\@.~\ref{fig1}(b). Thus, the discriminant number from Eq.~(\ref{eq:discriminant-number})
has definite signs within $K^{2}$, and behaves just like those in
a Brillouin torus.

Furthermore, as shown in Fig\@.~\ref{fig1}(c), within the region
$K^{2}$, the oriented loop $\text{\ensuremath{\Gamma}}$ can be deformed
to $\tilde{\Gamma}$ and further splitted into two 1D closed paths
denoted as $L_{1}$ and $-L_{2}$, due to topological robustness.
Consequently, the topological charge carried by EPs can be computed
by the difference between the discriminant number of the two paths
at different $k_{y}$ values, which is given by
\begin{equation}
v(k_{y})=\frac{i}{2\pi}\int_{-\pi}^{\pi}dk_{x}\cdot\partial_{k_{x}}\log\Delta(k_{x},k_{y}).\label{eq:ky-discriminant}
\end{equation}

\textit{\textcolor{blue}{Failure of fermion doubling theorem.}}\textit{\textemdash{}}
After specifying the local orientation within the region $K^{2}$,
the summation of topological charges for all EPs in $K^{2}$ can be
unambiguously computed as
\begin{equation}
\sum_{\mathbf{k}_{i}\in K^{2}}v(\mathbf{k}_{i})=\oint_{\partial K^{2}}d\mathbf{k}\cdot\nabla_{\mathbf{k}}\log\Delta(\mathbf{k}),
\end{equation}
as the integration paths can be continuously deformed to the Klein
bottle boundary $\partial K^{2}$, a process illustrated in Figs\@.~\ref{fig1}(d$_{1}$-d$_{3}$)
similar to the torus case~\citep{Chiu-NH_Double}.

Given that the boundary $\ensuremath{\partial K^{2}=abab^{-1}}$,
shown in Fig\@.~\ref{fig1}(d$_{3}$), the integration cancels out
on the two $b$ edges, while it adds up on the two $a$ edges. Hence,
we obtain
\begin{equation}
\sum_{\mathbf{k}_{i}\in K^{2}}v(\mathbf{k}_{i})=2\oint_{a}d\mathbf{k}\cdot\nabla_{\mathbf{k}}\log\Delta(\mathbf{k})\in2\mathbb{Z},\label{eq:sum-rule}
\end{equation}
since the integration along loop $a$ takes $\mathbb{Z}$ values. By
combining the above two equations, we arrive at Eq.~(\ref{eq:K2-rule})
in the introduction. Therefore, we can see that the fermion doubling
theorem for EPs fails in the Brillouin Klein bottle, as the summation equals
an even number.

It is crucial to address the role of the chosen region $K^2$ and its boundary $\partial K^2$. Because the Brillouin Klein bottle is a non-orientable manifold, a global orientation does not exist. Any calculation of topological charges, which relies on oriented path integrals, therefore requires the choice of a local orientation convention. Our definition of $K^2$ and its boundary corresponds to one such choice, often referred to as a ``cut.'' One might correctly argue that this choice is not unique; a different ``cut'' could be made, which would be equivalent to flipping the local orientation for a subset of the EPs. While such a change would alter the sign of individual charges $v_i \to -v_i$ for that subset, the difference between the old total charge sum ($S_{\text{old}}$) and the new one ($S_{\text{new}}$) would be $S_{\text{new}} - S_{\text{old}} = -2 \sum_j v_j$, where the sum is over the subset of EPs with flipped orientation. Since this difference is always an even integer, the \textit{parity} of the total charge sum is a robust topological invariant. Therefore, our central result---that the total charge is an even integer ($\sum v_i \in 2\mathbb{Z}$) and thus violates the fermion doubling theorem---is a fundamental property of the system, independent of the chosen convention for calculation.

Note that a similar proof can be done for Fermi points, as discussed
in the Supplemental Material (SM)~\citep{SuppInf}.

\textit{\textcolor{blue}{A concrete example violating the fermion
doubling theorem.}}\textit{\textemdash{}} Let us now investigate a
two-band model in the non-Hermitian Brillouin Klein bottle, characterized
by the Hamiltonian
\begin{equation}
\mathcal{H}_{\text{2}}(\mathbf{k})=(\alpha\cos k_{x}-\beta)\sigma_{1}+(\alpha\sin k_{x}-i\gamma\cos k_{y})\sigma_{2},\label{eq:two-band-model}
\end{equation}
where $\alpha$, $\beta$, and $\gamma$ are model parameters, and
$\sigma_{i}$'s are Pauli matrices. The model respects the symmetry
of Eq.~(\ref{eq:k-glide-sym}) with $U=\sigma_{1}$. We propose the
experimental realization of this model in the SM~\citep{SuppInf}.

\begin{figure}[t]
\includegraphics[width=0.95\columnwidth]{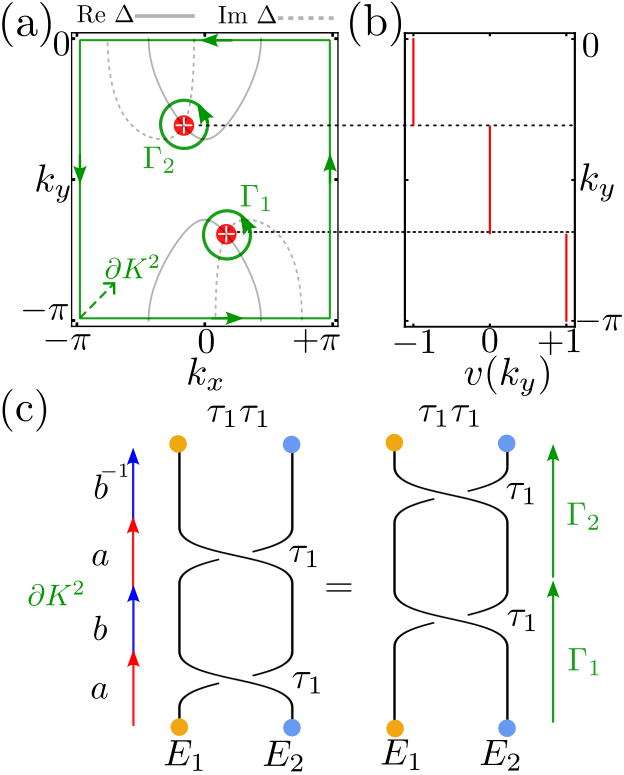}

\caption{(a) A two-band model where two EPs both having positive topological
charges. EPs are located at the position where $\text{Re}\,\Delta(\mathbf{k})=\text{Im}\,\Delta(\mathbf{k})=0$.
(b) The $k_{y}$-resolved topological invariants calculated using
Eq.~(\ref{eq:ky-discriminant}). (c) The braiding topology of EPs
in the Brillouin Klein Bottle. The parameters are $\alpha=0.5,$ $\beta=1.0-0.25i$,
and $\gamma=1.0$. The numerical calculations are given in the SM~\citep{SuppInf}.\label{fig2}}
\end{figure}

As depicted in Fig\@.~\ref{fig2}(a), there are two EPs located
at the position where both $\text{Re}\,\Delta(\mathbf{k})$ and $\text{Im}\,\Delta(\mathbf{k})$
vanish~\citep{SuppInf}. The two EPs both have a discriminant number
of +1, thereby violating the fermion doubling theorem. This is evident
by computing the $k_{y}$-resolved topological invariant $v(k_{y})$
of Eq.~(\ref{eq:ky-discriminant}). As shown in Fig\@.~\ref{fig2}(b),
$\ensuremath{v(k_{y})}$ both increases by +1 when crossing both EPs
by decreasing $k_{y}$, due to their identical topoloigcal charge
of $+1$. 

\textit{\textcolor{blue}{Braiding topology of EPs in Brillouin Klein
Bottles.}}\textit{\textemdash{}} Recently, it has been revealed that
EPs exhibit a richer braiding topology that cannot be fully captured
by the discriminant number~\citep{Zhi-Homotopical,Hu_ep_knot,Hu_Exceptional_Non-Abelian,Wojcik_eigenvalue,Wojcik_Homotopy,patil2022measuring,Lukas_PhysRevResearch}.
To explore this, we start by treating EPs as punctures in the Brillouin
Klein bottle $\mathsf{K}^{2}$~\citep{Boundary}. The set of a total
of $n$ EPs, denoted as $P=\{\mathbf{k}_{1},\mathbf{k}_{2},\ldots,\mathbf{k}_{n}\}$,
is isolated from $\mathsf{K}^{2}$ to ensure there is no eigenvalue
degeneracy on the $n$-punctured Klein bottle $\mathsf{K}^{2}-P$.
In this configuration, the $\ensuremath{N}$ complex eigenvalues $\{E_{1},\cdots,E_{N}\}$
are distinct and belong to the unordered configuration space $\text{UConf}_{N}(\mathbb{C})$~\citep{Zhi-Homotopical,Wojcik_eigenvalue}.
To reveal the braiding topology, we consider the set of homotopy classes
of maps from $\ensuremath{\mathsf{K}^{2}-P}$ to $\text{UConf}_{N}(\mathbb{C})$,
denoted by $[\mathsf{K}^{2}-P,\text{UConf}_{N}(\mathbb{C})]$. For
$\text{UConf}_{N}(\mathbb{C})$, only its fundamental group is non-trivial
and is equal to the braid group, $\pi_{1}(\text{UConf}_{N}(\mathbb{C}))=B_{N}$,
while all higher homotopy groups are trivial. Such a space is called
an Eilenberg-MacLane space of type $K(G,1)$, with $G=B_{N}$. An
important property of this space is that there exists a natural bijection~\citep{allen,Wojcik_eigenvalue}
\begin{equation}
[\mathsf{K}^{2}-P,\text{UConf}_{N}(\mathbb{C})]=\text{Hom}(\pi_{1}(\mathsf{K}^{2}-P),B_{N}).
\end{equation}
Hence, we need to compute the set $\text{Hom}(\pi_{1}(\mathsf{K}^{2}-P),B_{N})$
of the group homomorphisms from $\pi_{1}(\mathsf{K}^{2}-P)$ to $B_{N}$. 

The key difference from the Brillouin torus case is that the base
manifold is now the $n$-punctured Klein bottle $\mathsf{K}^{2}-P$.
Its fundamental group is $\pi_{1}(\mathsf{K}^{2}-P)=\{a,b,\Gamma_{1},\Gamma_{2},\cdots,\Gamma_{n}|abab^{-1}=\Gamma_{1}\Gamma_{2}\cdots\Gamma_{n}\}$,
as calculated in the SM~\citep{SuppInf}. Here, two generators are
from $\mathsf{K}^{2}$ (specifically, loops $a$ and $b$ on the edges)
and $n$ generators are from the loops surrounding the EPs, 
denoted as $\Gamma_{i}=\Gamma(\mathbf{k}_{i})$ around
the $i$th EP. There are $(n+2)$ generators constrained by the relation
\begin{equation}
abab^{-1}=\Gamma_{1}\Gamma_{2}\cdots\Gamma_{n},\label{eq:klein-relation}
\end{equation}
which can be understood as the continuous deformation of the loops
$\Gamma_{1},\cdots,\Gamma_{n}$ to the boundary of $\mathsf{K}^{2}$,
a process illustrated in Figs\@.~\ref{fig1}(d$_{1}$-d$_{3}$).
Thus, $\pi_{1}(\mathsf{K}^{2}-P)$ is a free group of $n+1$ generators
$\pi_{1}(\mathsf{K}^{2}-P)=*_{n+1}\mathbb{Z}$~\citep{Footnote}. 

Next, we compute the set of group homomorphisms $\text{Hom}(\pi_{1}(\mathsf{K}^{2}-P),B_{N})$.
A group homomorphism is a function $f:\pi_{1}(\mathsf{K}^{2}-P)\rightarrow B_{N}$,
which maps loops $\ensuremath{\gamma_{i}}$ in $\mathsf{K}^{2}-P$
to braid elements $b_{i}$ in the braid group $B_{N}$. By definition,
the group homomorphism preserves the group structure, i.e., $\ensuremath{f(\gamma_{1}\gamma_{2})=f(\gamma_{1})f(\gamma_{2})},$
and is determined by its values on the generators of $\pi_{1}(\mathsf{K}^{2}-P)$.
We denote the values of the homomorphism on the generators $\ensuremath{a}$
and $b$ as $f(a)=\mathsf{b}_{a}$ and $f(b)=\mathsf{b}_{b}$, respectively,
and on the generators $\Gamma_{i}$ as $f(\Gamma_{i})=\mathsf{b}_{i}$.
Furthermore, the homomorphism must preserve the relation given by
Eq.~(\ref{eq:klein-relation}). Thus, we obtain the braiding topology
as
\begin{align}
\text{Hom}(\pi_{1}(\mathsf{K}^{2}-P),B_{N})=\nonumber \\
\{\mathsf{b}_{a},\mathsf{b}_{b},\mathsf{b}_{1},\cdots,\mathsf{b}_{n}\in B_{N}| & \mathsf{b}_{1}\cdots\mathsf{b}_{n}=\mathsf{b}_{a}\mathsf{b}_{b}\mathsf{b}_{a}\mathsf{b}_{b}^{-1}\},\label{eq:braiding-topo}
\end{align}
which endorses a no-go theorem on the braiding patterns of EPs in
the Brillouin Klein bottle. That is, the composite braidings around all EPs
$\mathsf{b}_{1}\mathsf{b}_{2}\cdots\mathsf{b}_{n}$
must equal to the braiding $\mathsf{b}_{a}\mathsf{b}_{b}\mathsf{b}_{a}\mathsf{b}_{b}^{-1}$
along the Klein bottle boundary.

\begin{figure}[t]
\includegraphics[width=0.95\columnwidth]{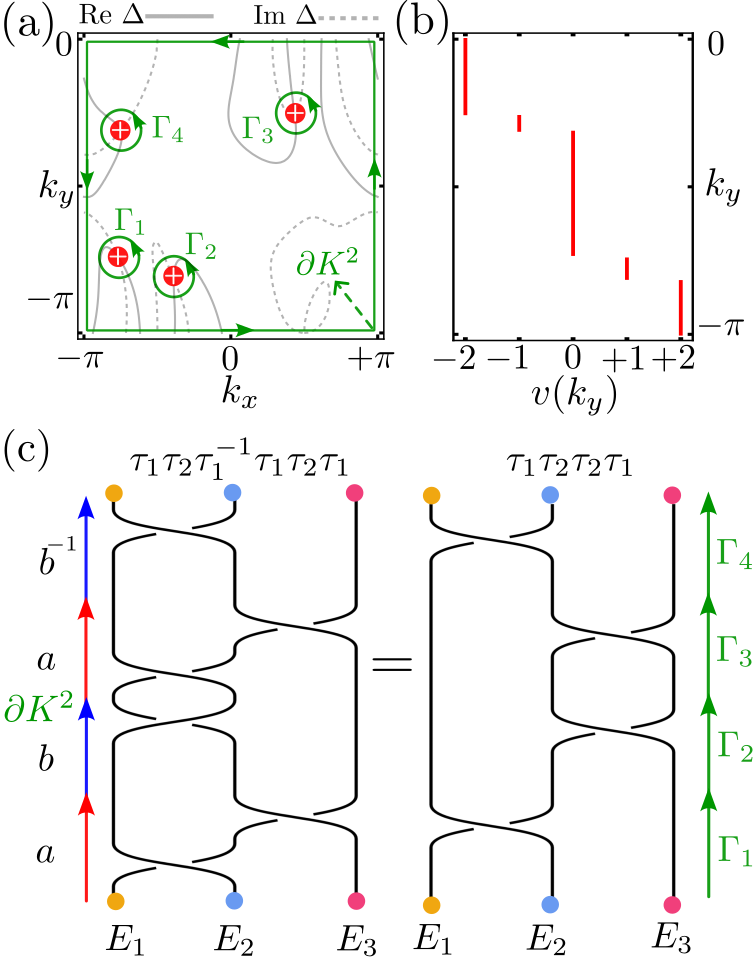}

\caption{Non-abelian braidings. (a) A three-band model where four EPs all having
positive topological charges. (b) The $k_{y}$-resolved topological
invariants calculated using Eq.~(\ref{eq:ky-discriminant}). (c)
The braiding of EPs in the Brillouin Klein bottle. The parameters
are $\alpha=2.0,$ $\beta=1.0$, $\gamma=1.0,$ $\delta=0.5$ and
$\epsilon=1.0$. The numerical calculations are given in the SM~\citep{SuppInf}.\label{fig3}}
\end{figure}

As shown in Fig\@.~\ref{fig2}(c) and Fig\@.~\ref{fig3}(c), the
braiding patterns can be represented by a sequence of over and under
crossings, after sorting the eigenenergies by their real parts. Specifically,
$\tau_{i}$ denotes where the $i$th eigenenergy crosses under the
$\ensuremath{(i+1)}$th eigenenergy, while $\tau_{i}^{-1}$ denotes
where it crosses over.

\textit{\textcolor{blue}{Revisiting the failure of fermion doubling
theorem.}}\textit{\textemdash{}} The failure of the fermion doubling
theorem can also be proven from the braiding topology. This is because
the discriminant number is determined by braid crossings: an over/under
crossing contributes $+1/-1$ to the discriminant numbers~\citep{Hu_ep_knot}.
As a result, the sum of the discriminant numbers of all EPs equals
those along the Klein bottle boundary, using the relation $\mathsf{b}_{1}\mathsf{b}_{2}\cdots\mathsf{b}_{n}=\mathsf{b}_{a}\mathsf{b}_{b}\mathsf{b}_{a}\mathsf{b}_{b}^{-1}$
in Eq.~(\ref{eq:braiding-topo}). Since the discriminant numbers
of $\mathsf{b}_{b}$ and $\mathsf{b}_{b}^{-1}$ cancel out, we obtain
the relation in Eq.~(\ref{eq:sum-rule}), where the sum equals twice
the discriminant numbers of $\mathsf{b}_{a}$.

Returning to the two-band model in Eq.~(\ref{eq:two-band-model}),
the braid group is $B_{2}$ for $N=2$, which is the abelian group
$\mathbb{Z}$ and all braid elements commute. As shown in Fig\@.~\ref{fig2}(c),
the braiding patterns on the loops $\ensuremath{\Gamma_{1}}$ and
$\Gamma_{2}$ are $\mathsf{b}_{1}=\mathsf{b}_{2}=\tau_{1}$, while
on the boundary, they are $\mathsf{b}_{a}=\tau_{1}$ and $\mathsf{b}_{b}=1$.
It can be verified that $\mathsf{b}_{1}\mathsf{b}_{2}=\mathsf{b}_{a}^{2}=\tau_{1}^{2}$,
in accordance with the braiding theory. The discriminant number for
all EPs sums up to $+2$, with each $\tau_{1}$ contributing $+1$,
further confirming the failure of the fermion doubling theorem.

\textit{\textcolor{blue}{Non-abelian braidings in Brillouin Klein
Bottle.}}\textit{\textemdash{}} The intriguing aspect of braiding
topology is that it becomes intrinsically non-abelian in multi-band
cases with $N\geq3$, which cannot be described by topological charges.
In this case, the $\ensuremath{\tau_{i}}'s$ satisfy the braid relations
$\tau_{i}\tau_{j}=\tau_{j}\tau_{i}$ for $|j-i|>1$ and $\tau_{i}\tau_{i+1}\tau_{i}=\tau_{i+1}\tau_{i}\tau_{i+1}$
for $1\leq i\leq N-1$. Consider a three-band model, whose Hamiltonian
is given by
\begin{equation}
\mathcal{H}_{\text{3}}(\mathbf{k})=\left(\begin{array}{ccc}
F(\mathbf{k}) & -1 & 0\\
-1 & 0 & -1\\
0 & -1 & G(\mathbf{k})
\end{array}\right),
\end{equation}
where $F(\mathbf{k})=\alpha\cos k_{x}+i\text{\ensuremath{\beta}}\sin k_{x}\cos k_{y}+i\epsilon$
and $G(\mathbf{k})=\gamma\cos2k_{y}+i\delta\sin2k_{y}-\epsilon$.
The model satisfies the symmetry of Eq.~(\ref{eq:k-glide-sym}) with
$U=\mathds{1}$. As shown in Fig\@.~\ref{fig3}(a), there are four
EPs all having topological charges of $\ensuremath{+1}$, thereby
violating the fermion doubling theorem. This can be verified by the
$k_{y}$-resolved topological invariant $v(k_{y})$ of Eq.~(\ref{eq:ky-discriminant}),
as shown in Fig\@.~\ref{fig3}(b).

As depicted in Fig\@.~\ref{fig3}(c), the braiding patterns on the
edges are $\mathsf{b}_{a}=\tau_{1}\tau_{2}$ and $\mathsf{b}_{b}=\tau_{1}^{-1}$.
They do not commute, i.e., $\mathsf{b}_{a}\mathsf{b}_{b}\neq\mathsf{b}_{b}\mathsf{b}_{a}$,
because $\mathsf{b}_{a}\mathsf{b}_{b}=\tau_{1}\tau_{2}\tau_{1}^{-1}$
is not equivalent to $\mathsf{b}_{b}\mathsf{b}_{a}=\tau_{2}$ using
the aforementioned braid relations. Thus, the braiding topology cannot
be reduced to $\mathsf{b}_{1}\mathsf{b}_{2}\cdots\mathsf{b}_{n}=\mathsf{b}_{a}\mathsf{b}_{b}\mathsf{b}_{a}\mathsf{b}_{b}^{-1}\neq\mathsf{b}_{a}^{2}$,
in contrast to the two-band case. On the other hand, the four EPs
have the braidings $\mathsf{b}_{1}=\tau_{1}$, $\mathsf{b}_{2}=\tau_{2}$,
$\mathsf{b}_{3}=\tau_{2}$, and $\mathsf{b}_{4}=\tau_{1}$ along the
loops $\Gamma_{1}$, $\Gamma_{2}$, $\Gamma_{3}$, and $\Gamma_{4}$,
respectively. It can be checked that $\mathsf{b}_{a}\mathsf{b}_{b}\mathsf{b}_{a}\mathsf{b}_{b}^{-1}=\mathsf{b}_{1}\mathsf{b}_{2}\mathsf{b}_{3}\mathsf{b}_{4}=\tau_{1}\tau_{2}\tau_{2}\tau_{1}$
as shown in Fig\@.~\ref{fig3}(c), in accordance with Eq.~(\ref{eq:braiding-topo}).
The non-commutative braiding relations highlight the necessity of
considering the full braid group structure to understand the topological
properties of EPs in the Brillouin Klein bottle.

\textit{\textcolor{blue}{Conclusions and discussions.}}\textit{\textemdash{}
}We have demonstrated that the well-established fermion doubling theorem
for EPs breaks down in the non-Hermitian Brillouin Klein bottle. Specifically,
we have proven that the sum of the total charges is an even number
rather than zero for the region without crossing tbe boundary. Moreover,
we have uncovered a novel braiding topology for EPs, particularly
the non-Abelian ones, which is distinct from that in the Brillouin
torus. While our study has primarily focused on gapless non-Hermitian
topologies, exploring gapped topological phases~\citep{Gong-topology,Kawabata-symmetry}
would be a promising direction for future research. Beyond bulk topologies,
it would also be valuable to investigate boundary effects, such as
non-Hermitian skin effects and topological boundary states~\citep{Wang-Amoeba,Nakamura-bbc},
in the Brillouin Klein bottle.

Our findings also pave the way for other intriguing research directions. For gapped phases, the non-orientable nature of the Brillouin Klein bottle would necessitate a new topological classification scheme beyond the standard ten-fold way classification. Another exciting frontier is the extension of our work to strongly correlated non-Hermitian systems. While a formidable challenge, investigating how many-body interactions affect the unique topological rules of the Brillouin Klein bottle, likely via a Green's function approach, promises to uncover even richer physics.

\vspace{0.5cm}

\begin{acknowledgements} 
The authors are grateful for the valuable discussions with Prof. Y. X. Zhao.
This work was supported by the Quantum Science
Center of Guangdong-Hong Kong-Macau Greater Bay Area, the NSFC/RGC
JRS grant (RGC Grant No. N\_HKU774/21, NSFC Grant No. 12161160315),
and the GRF of Hong Kong (Grant Nos. 17310622 and 17303023). W.B.R.
was supported by the RGC Postdoctoral Fellowship (Ref. No. PDFS2223-7S05).
\end{acknowledgements}

\bibliography{Reference}

\begin{thebibliography}{80}
\bibitem [{Bou()}]{Boundary}%
  \BibitemOpen
\bibfield  {journal} {  }\href@noop {} {}\bibinfo {note} {Note that a different
  font of $\mathsf{K}^{2}$ is used to denote the full Klein bottle, including
  its boundary}\BibitemShut {NoStop}%
\bibitem [{Foo()}]{Footnote}%
  \BibitemOpen
  \href@noop {} {}\bibinfo {note} {The fundamental group
  $\pi_{1}(\mathsf{K}^{2}-P)=*_{n+1}\mathbb{Z}$ as a free group of $n+1$
  generators can also be seen by the deformation retraction of
  $\mathsf{K}^{2}-P$: it reduces to the wedge product of $n+1$ circles
  $\vee_{n+1}S^{1}$, i.e., $n+1$ circles joined at a point. Here, the free
  group of $n+1$ generators is denoted by
  $*_{n+1}\mathbb{Z}=\underbrace{\mathbb{Z}*\mathbb{Z}*\cdots*\mathbb{Z}}_{n+1}$,
  and the wedge product is $\vee_{n+1}S^{1}=\underbrace{S^{1}\vee
  S^{1}\vee\cdots\vee S^{1}}_{n+1}$}\BibitemShut {NoStop}%
\end{thebibliography}

\end{document}